\documentclass[aps,pra,twocolumn,superscriptaddress,amsmath,amssymb]{revtex4}

\usepackage{amssymb}
\usepackage{bm}
\usepackage{epsfig}
\usepackage[ansinew]{inputenc}
\usepackage{graphicx}

\begin{document}

\title{Interplay of quantum and classical fluctuations near quantum
  critical points}

\author{M. A. Continentino}
\email{mucio@cbpf.br}
\affiliation{Centro Brasileiro de Pesquisas F\'{\i}sicas, Rua
  Dr. Xavier Sigaud 150, 22290-180, Rio de Janeiro, RJ, Brazil}

\date{\today}

\begin{abstract}
  For a system near a quantum critical point (QCP), above its lower critical
  dimension $d_L$, there is in general a critical line of second order
  phase transitions that separates the broken symmetry phase at finite
  temperatures from the disordered phase. The phase transitions along
  this line are governed by thermal critical exponents that are
  different from those associated with the quantum critical point.  We
  point out that, if the effective dimension of the QCP, $d_{eff}=d+z$
  ($d$ is the Euclidean dimension of the system and $z$ the dynamic
  quantum critical exponent) is above its upper critical dimension
  $d_C$, there is an intermingle of classical (thermal) and quantum
  critical fluctuations near the QCP. This is due to the breakdown of
  the generalized scaling relation $\psi=\nu z$ between the shift
  exponent $\psi$ of the critical line and the crossover exponent $\nu
  z$, for $d+z>d_C$ by a \textit{dangerous irrelevant
    interaction}. This phenomenon has clear experimental consequences,
  like the suppression of the amplitude of classical critical
  fluctuations near the line of finite temperature phase transitions
  as the critical temperature is reduced approaching the QCP.
\end{abstract}

\maketitle


\section{Introduction}

The modern theory of critical phenomena came to life four decades ago,
when Wilson presented renormalization-group tools honed to deal with
divergent free energies \cite{wilson71}. Soon applied to a number of
classical second-order phase transitions, those tools determined the
critical exponents associated with static properties, such as $\nu$,
associated with the divergence of the correlation length, or $\alpha$,
with the divergence of the specific heat, as well as the exponent $z$,
associated with critical slowing down and necessary to describe
dynamical properties.

Later, in the mid 1970's, a pair of developments extended the scope of
the renormalization-group approach to quantum Hamiltonians. The first
development started with two analyses of quantum models undergoing
phase transitions \cite{young, hertz}. Young considered the
$d$-dimensional Ising model subject to a transverse magnetic field
$h$ in the vicinity of the critical field $h_{c}$ that
destroys the magnetic order at zero temperature \cite{young}. He
showed that, at $T=0$, the model mimics the classical
$d+z$-dimensional Hamiltonian, where $z=1$ is the dynamic exponent
for the Ising model. At nonzero temperatures, by contrast, the
critical behavior is that of the $d$-dimensional model with
$h=0$. To the same general conclusions came Hertz \cite{hertz},
who considered a broader class of models, including ones with
$z>1$. Quantum fluctuations intertwine the static and dynamical
exponents; at the critical point, however, thermal fluctuations
overcome them.

The second line of development was equally important. By the end of
1974, its remarkable record of successes notwithstanding,
renormalization-group theory had covered but narrow regions of phase
diagrams, and some feared that it would never go beyond that.
Fortunately, such fears were unfounded. The formalism had a great
deal more power under its hood, which became visible when the Kondo
Hamiltonian was diagonalized and its physical properties, computed
\cite{wilson75}.  The Kondo model, a single impurity coupled to a
Fermi gas, undergoes no phase transition; nonetheless, when a
renormalization-group transformation was constructed, two fixed points
became apparent, and it was shown that the unusual physical properties
of the model reflect the crossover of the Hamiltonian from the high-
to the low-temperature fixed point \cite{wilson75}.

A decade later, when the same transformation was applied to the
two-impurity Kondo problem, a QCP was identified at the meeting point
of two competing tendencies: the coupling between each impurity and
the Fermi gas, which tends to screen the impurity moment, and the
antiferromagnetic interaction between the impurities, which tends to
lock the impurity moments into a singlet effectively decoupled from
the electron gas \cite{jvw}. The fixed point associated with the QCP
is analogous to the one discussed by Young \cite{young}; in contrast
with Ref.~\onlinecite{young}, however, Ref.~\onlinecite{jvw} was able
to extract the phase diagram for the two-impurity Kondo model from the
renormalization-group streamlines.

The insight derived from that work offered an opportunity to perfect
the then prevalent ideas concerning the phase diagram of the Kondo
lattice. This realized, in a collaboration with Japiassu and Troper
the present author envisioned a hypothetical renormalization-group
transformation, analogous to the one discussed in
Ref.~\onlinecite{jvw}, argued that it should have two fixed points---a
zero-temperature quantum fixed point and a nonzero temperature
classical one---and carried out a scaling analysis to discuss the
renormalization-group flow of the model Hamiltonian in their vicinity
\cite{mac0}.

Dependent only on basic assumptions, the scaling analysis proved to be
as powerful as it is general. It offers an enlightening view of phase
diagrams, one that pinpoints the origin of the singularities and
affords easy derivation of scaling relations. Examples will be
discussed below. Applied to the Kondo lattice, it showed that the
quantum fixed point is unstable against thermal excitations and
identified special renormalization-group flow lines stemming out of the
unstable fixed point with boundaries in the phase diagram. More
generally, the same construction provides insight and quantitative
information on the behavior of such strongly correlated systems as the
heavy-fermions and the high-$T_{c}$ compounds. Given the instability of
the fixed point, renormalization group flow lines are expected to
emanate from the QCP and carry information about it even to points
that are distant from the quantum phase transition in the phase
diagram. 

The recovery of that information is one of the exciting
problems in modern condensed-matter theory. This paper focuses that
problem in a restricted domain of the phase diagram, the immediate
vicinity of the QCP, and shows that, under conditions frequently found
in the laboratory, thermal fluctuations may be damped by quantum
fluctuations. Both the theoretical and the experimental facets of this
finding are examined.

\section{History}
\label{sec:1}

In the wake of Ref.~\onlinecite{mac0}, numerous studies of quantum
phase transitions have been reported
\cite{sachdev,mac,mac3,sim,qft1,qft2,si,flouquet,pairing,caldas}. Experimental
studies have focused their effects on the finite temperature behavior
of different physical quantities, sometimes even at unexpectedly high
temperatures. Quantum phase transitions occur in a large variety of
systems---metals, insulators, superconductors
\cite{sachdev,mac,si,pairing}---whenever a parameter can be tuned by
external means to drive a system in or out of a broken symmetry phase
at zero temperature. Disorder can also be used as a control parameter
\cite{saguia}. And the novel area of cold atom systems has opened
broad new perspectives, since the strength of the interactions, which
is generally fixed in condensed matter materials, can now be varied
and used to explore different types of phase diagrams close to a zero
temperature phase transition \cite{pairing}.

Theoretically, many techniques have been used to understand quantum
phase transitions, such as, simulations \cite{sim}, different types of
approximations \cite{qft1,moriya,pfeuty}, scaling theories
\cite{mac0,mac3} and the renormalization group, both in real
\cite{raimundo,pfeuty} and momentum space
\cite{hertz,millis,qft1}. These approaches yield the phase diagrams of
the relevant models, the behavior of physical quantities and the
universality class of quantum critical points. In many cases, theory
has been quite successful in explaining and even anticipating
experimental results \cite{sachdev,mac,mac0}. However there are
situations where the experimental results seem to imply a
dimensionality for the critical quantum fluctuations of the system
different from that expected from its crystalline structure.  This is
the case of the ubiquitous logarithmic term in the specific heat of
heavy fermions near an antiferromagnetic QCP. This logarithmic term
arises in a $d=2$, $z=2$ universality class, but the systems showing
this behavior in spite of anisotropic are clearly three dimensional
\cite{loh}.
 
As explained above, a distinctive feature of quantum phase transitions
is the inextricability of static and dynamics in the critical
phenomena \cite{young, hertz}. A clear manifestation of this entanglement
between space and time fluctuations is expressed by the quantum
hyperscaling relation \cite{mac3}, $2-\alpha=\nu(d+z)$ where $\alpha$
and $\nu$ are standard critical exponents characterizing the
non-analytic behavior of the free energy and the divergence of the
correlation length, respectively \cite{mac0}. The dimensionality of
the system $d$ appears in this relation modified by the dynamic
exponent $z$. It is through this relation and the scaling of
temperature near the QCP that the dynamic exponent enters in the
expression for the singular part of the free energy \cite{mac2,mac4}
\begin{equation}\label{free0} f_S \propto
|g|^{2-\alpha}F\left(\frac{T}{|g|^{\nu z}},
\frac{h}{|g|^{\beta+\gamma}}\right)
\end{equation} and ends up determining the critical behavior of static
thermodynamic quantities. In Eq.~(\ref{free0}), $h$ is the field
conjugate to the order parameter, $\beta$ its critical exponent and
$\gamma$ that of the susceptibility. The parameter $g$ measures the
distance to the quantum phase transition occurring at $g=0$. The
appearance of the dynamic exponent in thermodynamic quantities is a
unique feature of quantum phase transitions \cite{mac2,mac4,bjp}.  It
turns out that if the effective dimension $d_{eff}=d+z$ associated
with a QCP is larger than its upper critical dimension $d_C$, the
critical exponents describing the quantum critical behavior are
Gaussian, or mean field exponents \cite{hertz}.

\section{Classical and quantum critical fluctuations}

\begin{figure}[h]
\begin{center}
\includegraphics[width=1\linewidth, keepaspectratio]{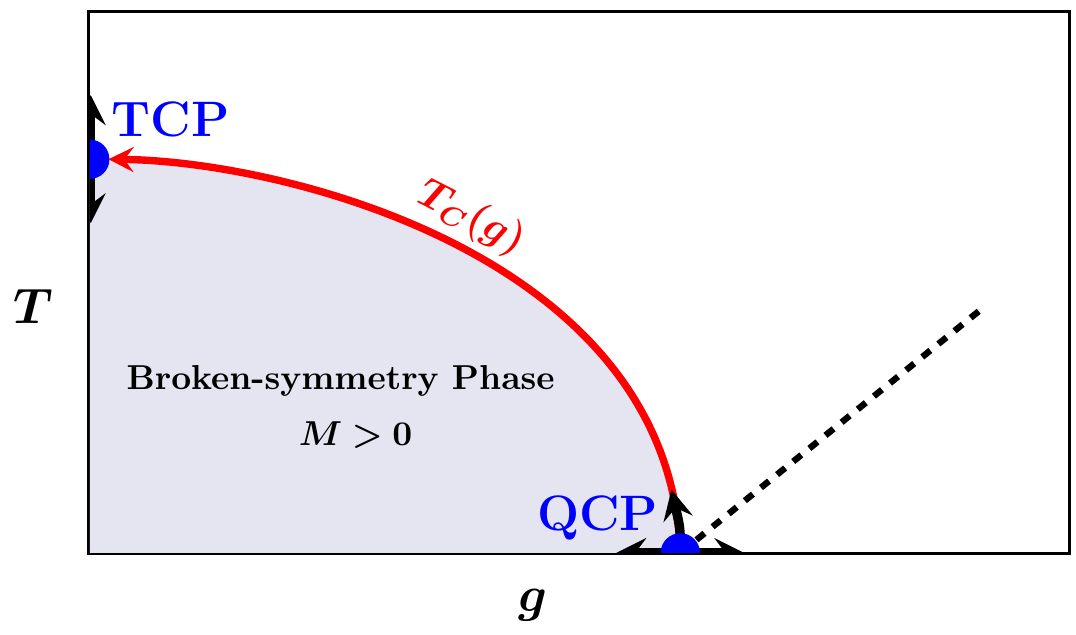}
\end{center}
\caption{(Color online) Generic phase diagram for a quantum system
above its lower critical dimension. As a function of the control
parameter $g$ there is a quantum phase transition at the quantum
critical point (QCP). There is also a critical line $T_{C}(g)$ of
finite temperature phase transitions which is governed by a thermal
critical point (TCP). The arrows show the flow of the renormalization
group equations.  } \label{fig1}
\end{figure}

In Fig.~\ref{fig1} we show a schematic phase diagram of a system
exhibiting quantum and thermal phase transitions. This phase diagram
is typical of quantum systems above their lower critical dimension
where the broken symmetry phase exists at finite temperatures. This
can be recognized, for example, as the phase diagram of the two
dimensional Ising model in a transverse field resulting from a real
space renormalization group approach \cite{raimundo}.

Classical and quantum critical phenomena are better described in the
language of the renormalization group, which associates critical
points with unstable fixed points of scaling transformations. The
phase diagram of Fig.~\ref{fig1} contains a fully unstable zero
temperature fixed point, the QCP that controls quantum criticality, and
a semi-unstable fixed point at finite temperatures, the thermal
critical point (TCP) governing the classical critical behavior along
the line of finite temperature phase transitions, $T_C(g)$. The arrows
in Fig.~\ref{fig1} show the flow of the renormalization-group
equations. Notice that the RG flow along the critical line is away
from the QCP and towards the TCP. 

In the figure, $g$ is a control parameter such that $g=0$ at the
QCP. In the specific case of the transverse Ising model, $g=h-h_C$,
with $h_C=(H/J)_C$ the critical ratio of the transverse field and the
Ising interaction.  The classical critical exponents are obtained from
an expansion of the renormalization-group equations in the
neighborhood of the thermal fixed point at TCP.  Since the RG flow
along the critical line runs towards it, the TCP determines the
critical behavior along the entire line. The QCP controls only the
zero-temperature phase transition. However, as pointed out before,
quantum fluctuations affect the finite temperature physical properties
making the properties of the QCP experimentally accessible, a point
that will be further discussed below \cite{mac1,mac}.

The phase diagram of Fig.~\ref{fig1} shows clearly a general feature of
critical phenomena in quantum systems above their lower critical
dimension $d_L$. Two universality classes coexist in
the problem, related to the thermal and quantum phase transitions
governed by the TCP and QCP, respectively.
 Expansions of the RG equations close to these fixed
points yield different sets of critical exponents.  In many cases of
interest, the dimensionality $d$ is smaller than the upper critical
dimensionality $d_{C}$, while $d+z>d_{C}$; in such cases the exponents
of the QCP are mean-field or Gaussian, while those describing the
singularities along the critical line $T_{C}(g)$ are the Wilson
exponents associated with the TCP, which are much harder to determine.

\section{Amplitude relations close to a quantum critical point}

For the sake of clarity we choose to present the arguments in this
section in the context of a real system showing both quantum and
thermal phase transitions. The insulating antiferromagnetic material
know as $DTN$ \cite{dtn}, with formula NiCl$_2$-4SC(NH$_2$)$_2$
presents two quantum phase transitions as a function of an external
magnetic field, which have been shown to be due to a Bose-Einstein
condensation of magnons. Also, it has a line of thermal phase
transitions separating the broken symmetry planar antiferromagnetic
phase from the disordered paramagnetic phase. 

\begin{figure}[h]
\begin{center}
\includegraphics[width=1\linewidth, keepaspectratio]{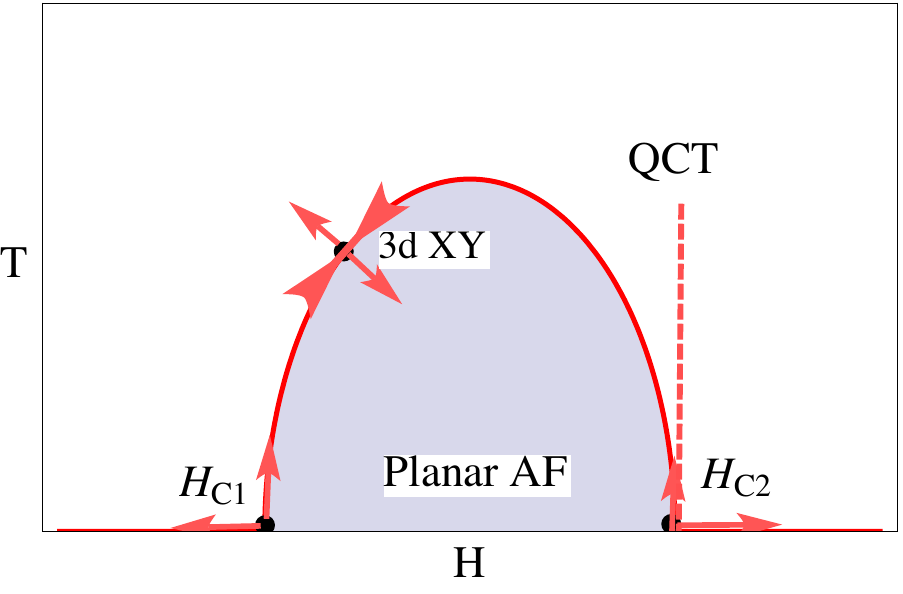}
\end{center}
\caption{(Color online) Schematic phase diagram of the
antiferromagnetic material NiCl$_2$-4SC(NH$_2$)$_2$. There are two
QCP's at $H_{C1}$ and $H_{C2}$ which are in the universality class of
the zero temperature density-driven three-dimensional Bose-Einstein condensation
with dynamic exponent $z=2$. The finite temperature transitions along
the critical line belong to the universality class of the classical
three-dimensional $XY$ model. The dashed line is one of the quantum critical
trajectories (QCT). } \label{fig2}
\end{figure}

A schematic temperature versus magnetic field phase diagram for this
system is shown in Fig.~\ref{fig2}. There are two QCP's at $H_{C1}$ and
$H_{C2}$ which are in the universality class of the density-driven
three-dimensional Bose-Einstein condensation, with dynamic exponent $z=2$
\cite{dtn}. On the other hand, the thermal phase transitions along the
critical frontier separating the planar antiferromagnetic from the
paramagnetic state are in the universality class of the classical three-dimensional
XY model \cite{3dxydtn}. This is associated with a hypothetical finite
temperature three-dimensional $XY$fixed point also shown in the figure together
with the expected RG flows.

Consider the zero temperature Bose-Einstein condensation phase
transition at the critical external magnetic field $H_{C1}$.  A similar
analysis can be carried out near $H_{C2}$. For small temperatures $T$,
sufficiently close to the QCP, the critical line is given by
$H_{C1}(T) = H_{C1}(1 + u T^{1/\psi})$ where $u$ is a constant related
to the magnon-magnon interaction \cite{dtnu}, and $\psi$ the {\it
shift exponent}.  Alternatively, the equation for this line close to
$H_{C1}$ can be written as $T_C(H) \propto |g|^{\psi}$, where
$g=H-H_{C1}$.  

For $d+z >4$ the interaction $u$ is a \textit{dangerous irrelevant}
quartic interaction \cite{mugnon} acting on the Gaussian fixed point
describing the QCP.  It is \textit{irrelevant} in the RG sense since
it scales to zero close to this fixed point. It is \textit{dangerously
  irrelevant} since it appears in the denominator of certain physical
quantities, as shown below. It is responsible in this case for
breaking down the generalized scaling relation \cite{millis} $\psi=\nu
z$.  

Figure~\ref{fig2} shows that the flow of the RG equations at finite
temperatures runs away from the QCP and towards the semi-stable three-dimensional
$XY$ fixed point (TCP). In the language of the RG, temperature is a
{\it relevant field} which scales away from the zero temperature
$QCP$. The flow is towards the semi-unstable TCP, which determines the
thermal critical exponents. The latter, which describe the thermal
phase transitions at $H_C(T)$ and are generally different from those
associated with the QCP's, will be denoted by a
\emph{tilde} to distinguish them from the critical exponents associated
with the QCP. We will thus refer to the quantum exponents
as $\{\alpha$, $\beta$, $\gamma$, $\nu$, $z$, $\cdots\}$ and to the
thermal critical exponents as $\{\tilde{\alpha}$, $\tilde{\beta}$,
$\tilde{\gamma}$, $\tilde{\nu}$, $\cdots\}$. 

In the first set we emphasize the special role of the dynamic quantum
critical exponent $z$, which appears in the expression for the free
energy, Eq.~(\ref{free0}). The classical counterpart $\tilde{z}$ has no
special significance, since it does not affect the critical behavior
of static thermodynamic quantities.

In each set, the critical exponents are bound by scaling relations,
for example, $\alpha+2 \beta +\gamma=2$ and
$\tilde{\alpha}+2\tilde{\beta}+\tilde{\gamma}=2$, which reduce the
number of independent critical exponents.  Particularly important is
the hyperscaling relation involving the dimensionality $d$ of the
system. In the classical case the hyperscaling relation is \cite{mac}
$2-\tilde{\alpha}=\tilde{\nu} d$. The quantum hyperscaling relation
\cite{mac3} in turn involves the dynamic quantum critical exponent and
is given by $2-\alpha=\nu(d+z)$. The appearance of an effective
dimension $d_{eff}=d+z$, has important consequences, for it decreases the
upper {\it Euclidean dimension} $d_C$ above which the system is
described by Gaussian or mean field exponents \cite{mac}. In DTN, since
the quantum phase transitions at $H_{C1}$ and $H_{C2}$ have $d_{eff}=
d+z=5 >d_C=4$, both are described by Gaussian or mean field exponents
\cite{mugnon}.

The above discussion appears to suggest that classical and quantum
phase transitions in general have separate descriptions not
interfering with each other. That this is not the case will be shown
below, on the basis of a generalized scaling approach \cite{pfeuty} (GSA)
extended for quantum phase transitions \cite{mac1}. More specifically,
we will see that, in the region of small nonzero temperatures in
the neighborhood of the QCP, if the system is above its lower critical
dimension $d_L$ and presents a line of second order thermal phase
transitions, the description of the quantum and thermal critical
behavior may involve, in each case, both quantum and thermal
exponents. 

In general this occurs if the quantum phase transition is above its
\textit{upper critical dimension}, i.e., if $d+z \ge d_C$ and
the generalized scaling relation $\psi=\nu z$ is violated due to a
dangerous irrelevant interaction \cite{millis}, which for the BEC
problem considered here is the magnon-magnon interaction
\cite{mugnon}. In this case, both quantum and thermal exponents are
necessary to describe the physical behavior along the quantum critical
trajectory, $g=0$, $T\rightarrow 0$.

For the Bose-Einstein condensation of magnons driven by an external
magnetic field, the transverse susceptibility $\chi_{_{\bot}}(H,T)$
plays the role of the order parameter susceptibility for this
transition. Close to the QCP at $H_{C0}$ ($H_{C0}=H_{C1}$ or
$H_{C0}=H_{C2}$) this susceptibility has the scaling form \cite{mac}
\begin{equation} \chi_{_{\bot}}(H,T) \propto |g(T)|^{-\gamma} Q\left(
\frac{T}{|g(T)|^{\nu z}}\right),
\end{equation} where $g(T)=|H-H_C(T)|/H_{C0}$ is the normalized
distance to the critical line $g(T_C)=0$ and
$H_C(T)=H_{C0}(1+uT^{1/\psi})$  in the phase diagram.

Continuity imposes the following asymptotic behaviors of the
scaling function $Q(t)$ ($t=T/|g(T)|^{\nu z}$) \cite{gsa,mac}:
\begin{itemize}
\item $Q(t=0) =$ constant. This guarantees that at zero temperature we
obtain the correct expression for the quantum critical behavior of the
transverse susceptibility, $\chi_{_{\bot}}(H,0) \propto
|g|^{-\gamma}$, such that, the order parameter susceptibility diverges
at zero temperature with the quantum critical exponent $\gamma$.
\item $Q\left( t \rightarrow \infty \right) \propto t^y$. The exponent
$y=(\tilde{\gamma}-\gamma)/\nu z$ yields the correct thermal critical
behavior of the transverse susceptibility close to the critical line
$g(T)=0$, i.e., $\chi_{_{\bot}}(H,T) \propto
A_{\bot}(T)|g(T)|^{-\tilde{\gamma}}$ as $T \rightarrow T_C(H)$.
\end{itemize} The amplitude $A_{\bot} (T) \propto
T^{(\tilde{\gamma}-\gamma)/\nu z}$ is determined by both classical and
quantum critical exponents. Then, close to the critical line, the order
parameter susceptibility diverges as
\begin{equation} \label{tres} \chi_{_{\bot}} \propto A_{\bot}(T)\,
|g(T)|^{-\tilde{\gamma}}
\end{equation} with the correct thermal critical exponent
$\tilde{\gamma}$.

We now consider a special trajectory in the phase diagram of
Fig.~\ref{fig2}: we let the system \textit{sit} at the QCP and lower
the temperature. Along this \textit{quantum critical trajectory}
(QCT), $g(0)=0$, i.e., for $H=H_C(T=0)=H_{C0}$ and $T\rightarrow 0$,
with help of the expression for the amplitude $A_{\bot}(T)$, we find
from Eq.~(\ref{tres}) that the transverse susceptibility diverges as,
\begin{equation}\label{chi} \chi_{_{\bot}}(H_{C0},T) \propto
T^{\frac{\tilde{\gamma}-\gamma}{\nu z}}
(uT^{\frac{1}{\psi}})^{-\tilde{\gamma}}.
\end{equation} 
Then, it follows that for $d+z>d_C$ in the presence of a
dangerous irrelevant interaction $u$ breaking down
the generalized scaling relation, such that $\psi \neq \nu z$, both
thermal and quantum critical exponents are required to give the
correct critical behavior along the QCT. Notice however, that for
$\psi=\nu z$, the exponent $\tilde{\gamma}$ cancels out in the above
expression and the divergence of the order parameter susceptibility,
$\chi_{_{\bot}}(H_{C0},T) \propto T^{-\gamma/\nu z}$ is governed only by
the critical exponents associated with the QCP. The temperature dependence of this result coincides
with that of a \textit{purely Gaussian} ($u=0$) theory, in which the
quartic interaction $u$ is neglected \cite{mac}. The equality between
the crossover and the shift exponents $\nu z=\psi$, i.e., the
generalized scaling relation is expected to hold for $d+z<d_C$.

For the correlation length, a similar analysis yields
\begin{equation}\label{xi} \xi \propto A_{L}(T) |g(T)|^{-\tilde{\nu}}
\end{equation} with $A_L(T) = T^{(\tilde{\nu}-\nu)/\nu z}$.  Along the
quantum critical trajectory,
\begin{equation}\label{QCTxi}
 \xi(H_{C0},T) \propto T^{\frac{\tilde{\nu}-\nu}{\nu z}
 }(uT^{\frac{1}{\psi}})^{-\tilde{\nu}},
\end{equation} and in general we find the same interference between
classical and quantum critical behavior for $\psi \neq \nu z$. Again,
for $\psi=\nu z$ the temperature dependence of $\xi(H_{C0},T)$
reduces to that of the purely Gaussian case, $\xi(H_{C0},T) \propto
T^{-1/z}$. Notice in Eqs.~(\ref{chi})~and (\ref{QCTxi}) that $u$ appears
in the denominator, justifying its dangerous irrelevant
character. Equations (\ref{chi})~and (\ref{xi}) and their respective
amplitudes give rise to a new effect, namely the \textit{quantum
suppression of classical fluctuations}, if the thermal exponents are
non-Gaussian.

The specific heat requires a more careful analysis since it vanishes
in the zero temperature axis. Its scaling behavior is obtained from
the scaling expression for the singular part of the free energy close
to the critical line which is given by,
\begin{equation}\label{free} f_S \propto
T^{\frac{\tilde{\alpha}-\alpha}{\nu z}}|g(T)|^{2-\tilde{\alpha}}.
\end{equation} The most singular contribution for the specific heat
close to this line is given by,
\begin{equation}\label{c} C/T \propto
T^{\frac{\tilde{\alpha}-\alpha}{\nu
z}}|g(T)|^{-\tilde{\alpha}}\left(g^{\prime}(T)\right)^2
\end{equation} where $g^{\prime}(T)=\partial g/\partial
T=-(uH_{C0}/\psi)T^{1/\psi -1}$. Along the QCT, if
 $\psi=\nu z$, the temperature dependence of this contribution
 coincides with that of the purely Gaussian result ( $u=0$)
 \cite{mac}, namely $C/T \propto T^{(d-z)/z}$.


 We have used in this section a generalized scaling approach to show
 that for $\psi \ne \nu z$, the critical behavior of physical
 quantities in the neighborhood of a QCP from which emanates a line of
 second order thermal phase transitions depends on both quantum and
 thermal exponents. Even along the quantum critical trajectory, $g=0$,
 $T \rightarrow 0$, the classical and quantum critical behaviors are
 intermingled, provided only that $\psi \neq \nu z$.

\section{Gaussian approximations for the classical critical behavior}

The full problem schematized in Fig.~\ref{fig1} is of very difficult
solution. The real space RG in general gives global phase diagrams and
all the relevant fixed points in the problem \cite{raimundo}. However,
in this approach, it is hard to obtain controlled expansions that
yield the correct critical exponents associated with these different
fixed points. And for electronic systems the real space RG has only
very limited success \cite{mac}. Momentum space RG \cite{millis} and
self-consistent approximations \cite{moriya} on the other hand give a
better treatment of strongly correlated electronic systems. The former
\cite{millis} describes correctly the unstable fixed point associated
with the QCP; unfortunately it is limited to the neighborhood of the
fixed point.  Although it describes the correct temperature runaway
flow away from the QCP, it fails to yield the thermal Wilson fixed
point to which the flow along the critical line is attracted.
\begin{figure}[h]
\begin{center}
\includegraphics[width=1\linewidth, keepaspectratio]{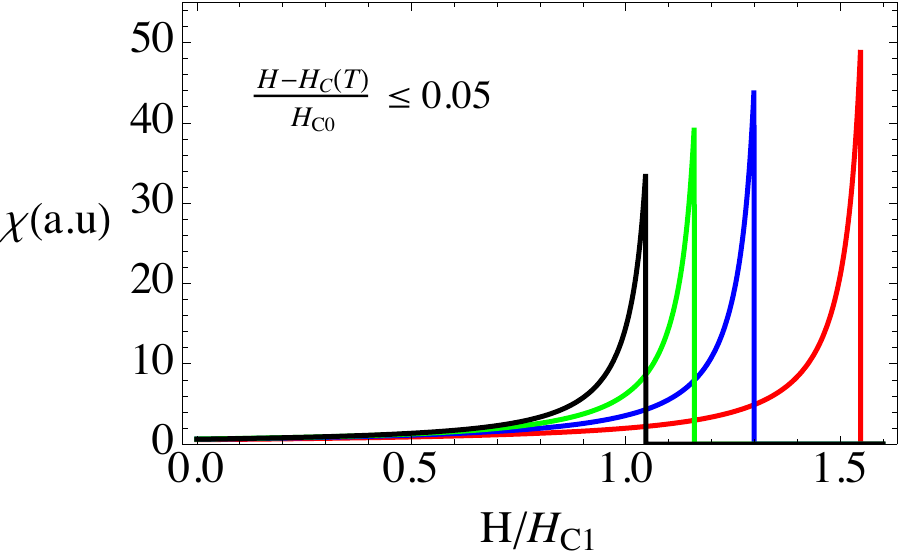}
\end{center}
\caption{(Color online) The order parameter susceptibility close to
the critical line $H_C(T)$ for fixed temperatures assuming that the
thermal exponents are those of the three-dimensional $XY$ model. The curves have
increasing temperature from left to right and extend to
$(H-H_C(T))/H_{C0}=5 \times 10^{-2}$. See Fig.~\ref{fig4} for
comparison. } \label{fig3}
\end{figure}
\begin{figure}[h]
\begin{center}
\includegraphics[width=1\linewidth, keepaspectratio]{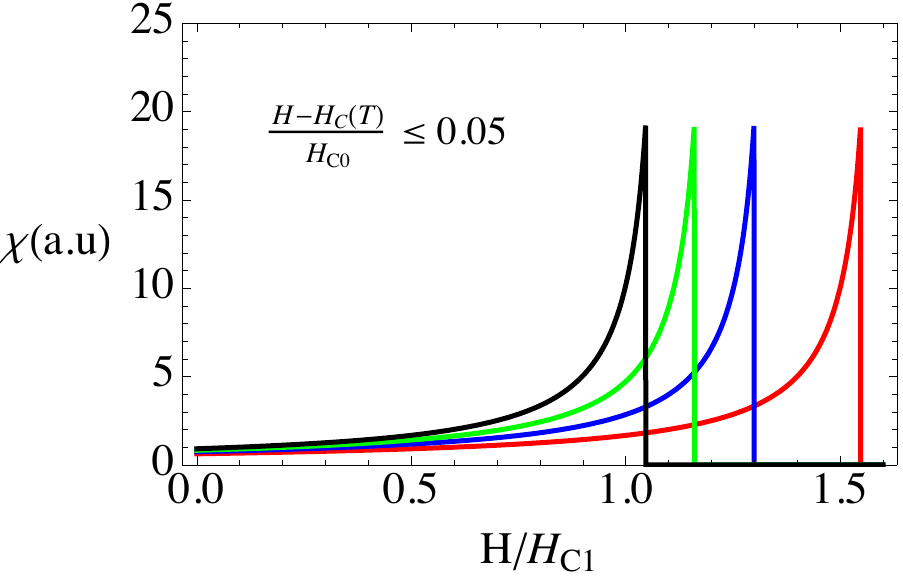}
\end{center}
\caption{(Color online) The order parameter susceptibility close to
the critical line $H_C(T)$ for fixed temperatures assuming Gaussian
exponents for the thermal phase transitions. The curves have
increasing temperature from left to right and extend to
$(H-H_C(T))/H_{C0}=5 \times 10^{-2}$.  } \label{fig4}
\end{figure} The self-consistent approach \cite{moriya} is a Gaussian
theory which yields the correct quantum critical exponents if the QCP
effective dimension is larger than the upper critical
dimension. However, it is inadequate to treat the thermal phase
transitions and yields Gaussian thermal exponents. In contrast with
the limitations of other approaches, the GSA emerges as the
perfect tool to understand the subtleties of quantum phase
transitions. It is particularly useful to explore the intermingle of
classical and quantum fluctuations near a QCP.

Both the momentum space RG and the self-consistent approach to quantum
phase transitions yield the shift exponent $\psi=z/(d+z-2)$,
showing clearly the breakdown of the scaling relation $\psi=\nu z$ for
$d+z\ge d_C$. The momentum space RG shows that this breakdown is due
to a dangerous irrelevant quartic interaction acting on the Gaussian
QCP. In the case of DTN, this is the magnon-magnon interaction $u$.

In this section we show that the results of the momentum space RG and
of the self-consistent approach for the critical behavior along the
quantum critical trajectory are recovered when we assume that the
\textit{thermal exponents} take thermal Gaussian values.  

We are interested in the case $d+z \ge d_C$, i.~e., in Gaussian \textit{quantum
critical exponents}.  Consider Eqs.~(\ref{chi}) and
(\ref{QCTxi}) for the order parameter susceptibility and correlation
length, respectively. Along the quantum critical trajectory, taking
$\tilde{\gamma}=\gamma$ and $\tilde{\nu}=\nu$ we get for the order parameter susceptibility,
\begin{equation}\label{chiG} \chi_{_{\bot}} \propto
(uT^{\frac{1}{\psi}})^{-\gamma}
\end{equation} and for the correlation length,
\begin{equation}\label{xiG} \xi \propto(uT^{\frac{1}{\psi}})^{-\nu},
\end{equation} which are the results from the momentum space RG and
self-consistent approach. These results are clearly distinct from
those given in Eqs.~(\ref{chi}) and (\ref{QCTxi}), respectively.

More importantly, the present analysis shows that substitution of
Gaussian values for the thermal (\textit{tilde})
critical exponents leads to temperature independent amplitudes,
since the quantum exponents are Gaussian.  This follows immediately from
Eqs.~(\ref{tres}) and (\ref{xi}) and the corresponding
amplitudes. Thus, if the thermal exponents turn out to be Gaussian,
there will be no suppression of classical fluctuations in the vicinity
of the QCP. This is
illustrated in Figs.~\ref{fig3} and \ref{fig4}.

The analysis for the specific heat is much more subtle. Differently
from other critical exponents, the quantum and thermal Gaussian
exponents $\alpha$ and $\tilde{\alpha}$ are different and determined
by the corresponding hyperscaling relations. For example, they yield
$(\tilde{\alpha}-\alpha)/\nu z=1$. We find that the most singular
contribution to the specific heat arising from the free energy,
Eq.~(\ref{free}), assuming that the \textit{tilde} exponents are
Gaussian thermal exponents is given by,
\begin{equation}\label{istoai} C/T \propto u^2
T^{\frac{2}{\psi}-1}|g(T)|^{-\tilde{\alpha}}.
\end{equation} Notice that in this case classical fluctuations are
suppressed even with Gaussian thermal exponents ($\psi < 2$).  Along
the QCT, using $\tilde{\alpha}=1/2$ in $d=3$, we find that
\begin{equation}\label{istoai2} C/T \propto u^{3/2} T^{\frac{3}{2\psi}
-1}.
\end{equation} Differently from Eqs.~(\ref{chiG}) and (\ref{xiG}) for the
susceptibility and correlation length, the quartic interaction $u$
appears in this case in the numerator.  Eq.~(\ref{istoai2}) is the
expected result for the specific heat along the QCT for $d=3$ assuming
that thermal exponents are Gaussian and taking into account the
dangerous irrelevant interaction $u$. However, the purely Gaussian 
result \cite{mac}, which neglects the quartic
interaction $u$ ( $u=0$), namely, $C/T \propto T^{(d-z)/z}$ is in
general more singular as $T\rightarrow 0$ then the contribution above, Eq.~(\ref{istoai2}). 
For the specific heat,
whether we consider the purely Gaussian result, which neglects the
interaction $u$, or Eq.~(\ref{istoai2}), which takes $u$ into account,
we find that the most singular contribution determines the physical
behavior along the QCT. This follows from Eqs.~(\ref{istoai}) and
(\ref{istoai2}), the right hand sides of which vanish for $u=0$ and
can be safely neglected with respect to the purely Gaussian ($u=0$)
contribution for the specific heat along the QCT. This is not the case
for the susceptibility and the correlation length, which diverge for
$u=0$. Finally, we point out that the same temperature dependence of
the pure Gaussian theory ($u=0$) for $C/T$ is obtained from
Eq.~(\ref{istoai2}), if we let $\psi=\nu z$ on the right-hand side.

\section{Bose-Einstein condensation in magnetic systems}

One consequence of the difference between thermal and quantum critical
exponents, as we have seen, is the suppression of the amplitude of
critical fluctuations along the critical line as the critical
temperature is reduced in the vicinity of a QCP. Besides, even along
the quantum critical trajectory, the thermal exponents appear in the
expressions for the critical behavior of different quantities. These
are strong results with clear experimental consequences. In this
section we discuss the implication of these results for magnetic
systems presenting a Bose-Einstein condensation of magnons as a
function of the magnetic field.

This is the case of DTN, which has a phase diagram \cite{dtn} similar
to that shown schematically in Fig.~\ref{fig2}. As pointed out before,
at $H_{C1}$ and $H_{C2}$, the quantum phase transitions of this three
dimensional system are in the universality class of the
density-driven Bose-Einstein condensation, with dynamic exponent
$z=2$. Since the effective dimensions of the QCP's at these critical
fields are $d_{eff}=3+2=5$ which is larger than $d_C=4$, the quantum
phase transitions are described by Gaussian or mean-field exponents.
In particular $\nu=1/2$, such that, $\nu z=1$. 

The shift exponent of the critical lines emanating from the QCP's is
given by $\psi =z/(d+z-2)=2/3$, as results from the momentum-space RG
analysis in Ref.~\onlinecite{millis}. This is clearly distinct from
the crossover exponent $\nu z=z/2=1$. The dangerously irrelevant
magnon-magnon interaction $u$ is responsible for the breakdown of the
scaling relation $\psi=\nu z$ in this case \cite{mugnon}. The
exponents we have just obtained describe well the critical behavior;
in particular, $\psi=2/3$ fits very well the critical lines close to
the critical fields $H_{C1}$ and $H_{C2}$
\cite{dtn,3dxydtn,dtnu,massren}. 

By contrast, the thermal or {\it tilde} exponents describing the
thermal phase transitions along the finite temperature critical line
are those of the three-dimensional $XY$ model. These Wilson critical
exponents are well known \cite{3dxytheo,3dxyexp}. They have been
obtained by different methods, from an RG $\epsilon$ expansion or
numerical methods \cite{3dxytheo}, and verified experimentally
\cite{3dxyexp}. These thermal exponents are clearly distinguishable
from the Gaussian ones associated with the QCP's. The specific heat
shows a clear $\lambda$ anomaly at the critical line associated with a
critical exponent $\tilde{\alpha} \approx 0$. We therefore have an 
opportunity to test the results in the previous sections.

Taking into account that the thermal critical exponents are different
from the Gaussian exponents associated with the QCP's, we obtain a
divergent order parameter susceptibility along the quantum critical
trajectory, $\chi_{_{\bot}}(H_{C0},T) \propto u^{-\tilde{\gamma}/\psi
}T^{-(1+\tilde{\gamma}/2)}$, where we used the Gaussian value for the
susceptibility exponent of the QCP, $\gamma=1$. Since $\tilde{\gamma}
\approx 1.32$ \cite{3dxytheo,3dxyexp} for the three-dimensional $XY$ model, this
yields a contribution more singular than that of the momentum space RG
and self-consistent approach, namely $\chi_{\bot}(H_{C0},T) \propto
(uT)^{-\gamma/\psi}=(uT)^{-3/2}$.

For the specific heat we find, $C/T \propto u^2 T^{3/2}\log T$ along
the quantum critical trajectory, $g(T=0)=0$, $T \rightarrow 0$. We used
$\tilde{\alpha} \approx 0$ for the specific heat exponent of the three-dimensional
$XY$ model \cite{3dxytheo,3dxyexp}, which implies a logarithmic
dependence in this case. However, this term is less singular then the purely Gaussian,
order zero in $u$ contribution to the specific heat, namely, $C/T
\propto T^{(d-z)/z} =T^{1/2}$.

Let us now investigate the behavior of the thermodynamic quantities at
the critical line for sufficiently low temperatures near the QCP's. The
influence of the quantum exponents due to the proximity of the quantum
phase transition is contained in the expressions for the
amplitude. Since the quantum critical exponents are Gaussian, we
obtain for the thermal critical behavior of the order parameter
susceptibility,
\begin{equation}\label{quibot} \chi_{_{\bot}}(H,T) \propto
T^{\tilde{\gamma}-1} \left| \frac{H-H_C(T)}{H_{C0}}
\right|^{-\tilde{\gamma}}
\end{equation} and for the specific heat,
\begin{equation}\label{cexp} C/T \propto u^2 T^{\frac{3}{2} +
\tilde{\alpha}}\left| \frac{H-H_C(T)}{H_{C0}}
\right|^{-\tilde{\alpha}}
\end{equation} where for the three-dimensional $XY$ model,
$\tilde{\gamma} \approx 1.32$ and $\tilde{\alpha} \approx 0$ implying
a lambda type anomaly for the specific heat \cite{3dxytheo,3dxyexp}
along the critical line. For completeness, we quote other critical
exponents for the three-dimensional $XY$ model, $\tilde{\nu} \approx 0.67$ and
$\tilde{\beta} \approx 0.34$ \cite{3dxytheo,3dxyexp}.

For the analysis of the experimental data on the critical behavior
along the critical line we can adopt the following procedures. Suppose
that, to approach the critical line in Fig.~\ref{fig2}, we decrease the
temperature for a fixed magnetic field $H_1 \gtrsim H_{C0}$.  We then
stop our measurements, say of the order parameter susceptibility, at a
given arbitrary temperature $T_1$, so distant from the thermal phase
transition that $[T_1-T_C(H_1)]/H_{C0}=\delta$. 

Next, we repeat the measurements for another fixed magnetic field $H_2
\gtrsim H_1$, stopping at another temperature $T_2$ so distant
from the critical line that $[T_2-T_C(H_2)]/H_{C0}=\delta$. When
this procedure is repeated for several magnetic fields, we find from
Eq.~(\ref{quibot}) that
the susceptibility at the temperatures $T_i$ such that
$[T_i-T_C(H_i)]/H_{C0}=\delta$ is given by, $\chi_{\bot}(T_i) \propto
T_i^{\tilde{\gamma}-1} \delta$ where $\delta$ is a fixed known constant. In this way using the values of 
$\chi_{\bot}(T_i)$ we can determine the exponent $\tilde{\gamma}$.  

Alternatively, we could increase the magnetic field at a fixed
temperature $T_1$. We would then stop the measurements at a field $H_1$ so
distant from the thermal phase transition that
$[H_1-H_C(T_1)]/H_{C0}=\delta$. We would repeat the measurement for another
fixed temperature $T_2$ and stop at a distance from the critical line,
such that $[H_2-H_C(T_2)]/H_{C0}=\delta$. The susceptibility at the
fields such that $[H_i-H_C(T_i)]/H_{C0}=\delta$ would then be given by
$\chi_{\bot} \propto T_i^{\tilde{\gamma}-1} \delta$ where $\delta$ is
a known constant.

Since, in practice, the susceptibility never diverges, we can take the
temperatures $T_i$ as the transition temperatures themselves.
In this case we can also work with the amplitudes given
directly in terms of the magnetic field. Since $T_i
=\left[(H_i-H_{C0})/uH_{C0}\right]^{\psi}$, substitution on the
right-hand side of Eq.~(\ref{quibot}) leads to $\chi_{\bot}(H_i) \propto
\left[(H_i-H_{C0})/uH_{C0}\right]^{\psi (\tilde{\gamma}-1)}$. The
envelope of the peaks in Fig.~\ref{fig3} is then the amplitude
function.

\begin{figure}[h]
\begin{center}
\includegraphics[width=1\linewidth, keepaspectratio]{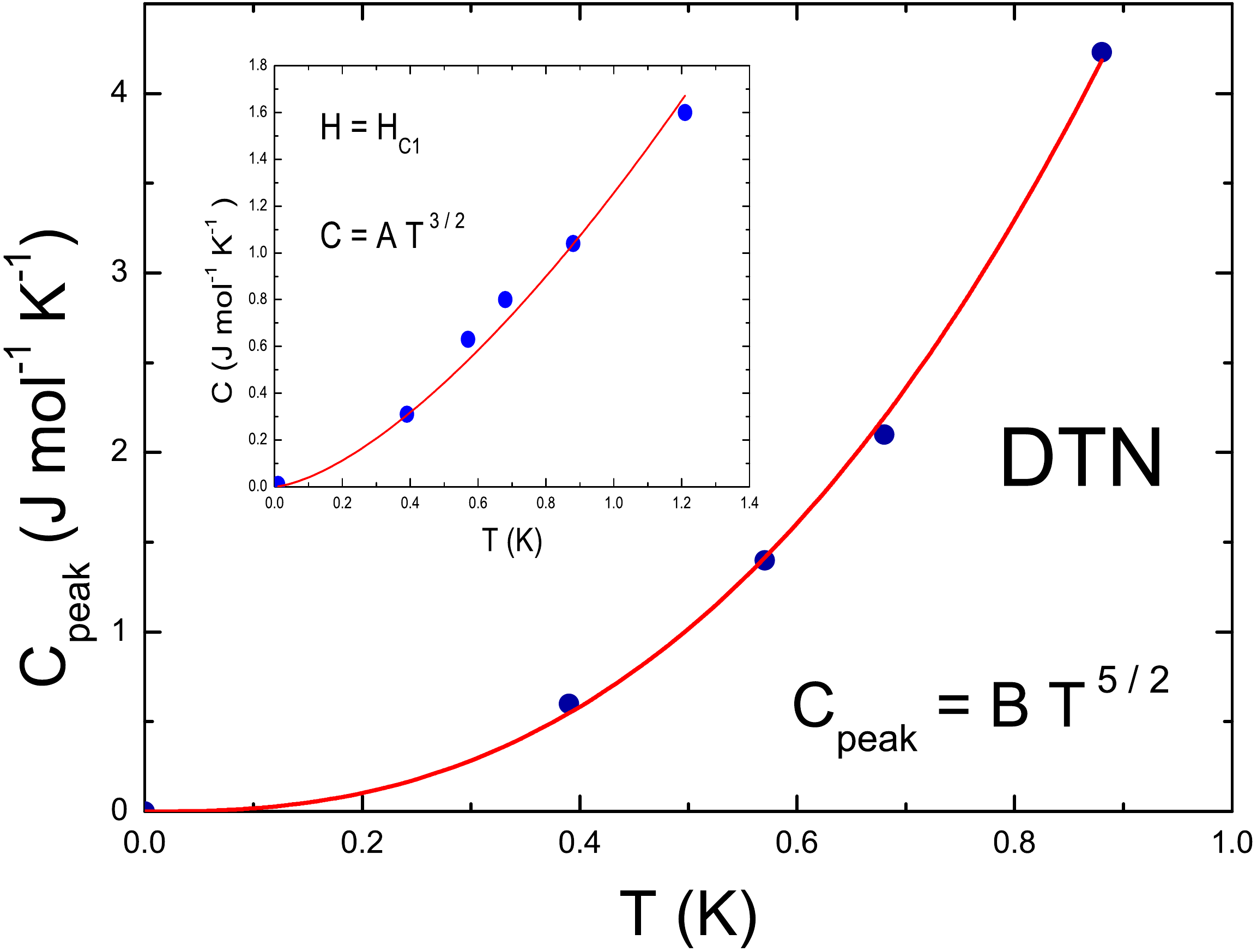}
\end{center}
\caption{(Color online) The amplitude of the specific heat peaks at
  the finite temperature phase transitions into the planar
  antiferromagnetic order or Bose condensed phase of DTN. The curve is
  a plot of $C \propto T^{5/2}$ obtained using $\tilde{\alpha} \approx
  0$ as appropriate for the three-dimensional $XY$ model. The inset
  shows the specific heat at the QCP at $H_{C1}$.  Points are data from
  Ref.~\onlinecite{nhmfl}.} \label{fig5}
\end{figure}

We have carried out the procedure outlined above for the DTN system
\cite{dtn} and also for $Sr_3Cr_2O_8$ \cite{aczel} which is another
magnetic system candidate for BE condensation \cite{aczel}. The
results for DTN are shown in Fig.~\ref{fig5} using the data near $H_{C1}$ from
Ref.~\onlinecite{nhmfl}. The figure shows the specific heat at the
peak temperatures, i.e., at the temperatures of the thermal phase
transitions for different magnetic fields. The line through the points
is $C \propto T^{5/2}$, as expected from Eq.~(\ref{cexp}) for
$\tilde{\alpha}=0$. 

\begin{figure}[h]
\begin{center}
\includegraphics[width=1\linewidth, keepaspectratio]{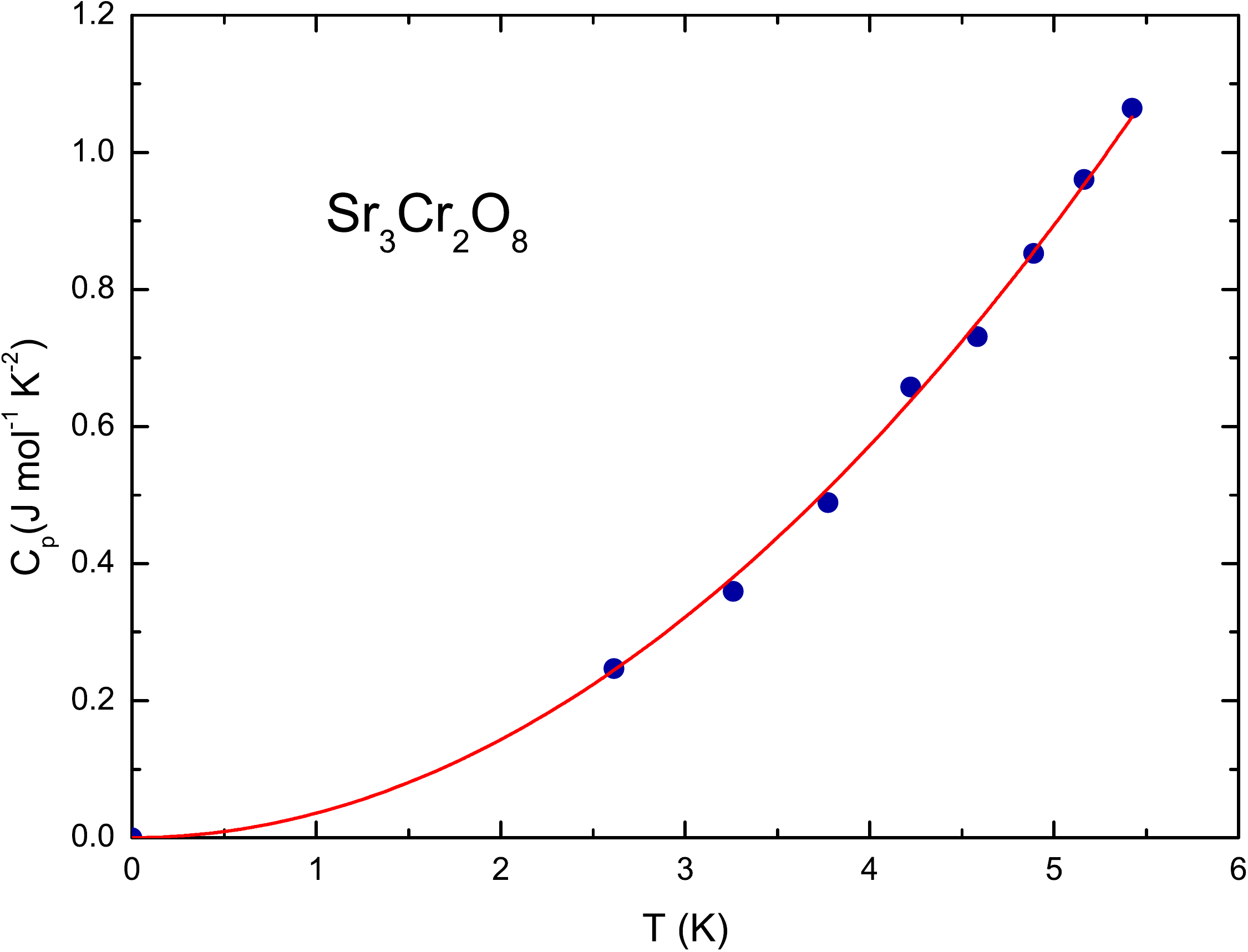}
\end{center}
\caption{(Color online) The amplitude of the specific heat peaks at
  the finite temperature phase transitions in $Sr_3Cr_2O_8$. The curve
  is a plot of $C \propto T^{3}$ with $\tilde{\alpha}
  \approx 1/2$, as appropriate for the three-dimensional Gaussian
  model. Points are data from
  Ref.~\onlinecite{aczel}. } \label{fig6}
\end{figure}

The fully Gaussian results, i.e., results obtained with thermal
gaussian exponents, give a stronger suppression of the specific heat
fluctuations namely, $C \propto T^{3}$. This seems to be the case for
$Sr_{3}Cr_2O_8$, which is shown in Fig.~\ref{fig6}; here, a $T^3$ law
fits very well the amplitude of the specific heat in
Ref.~\onlinecite{aczel} for various magnetic fields close to the
critical field. In this case, however, the $T^{3}$ dependence raises a
doubt: it is not clear whether we are observing critical fluctuations
or measuring the phonon contribution for the specific heat.  

In any case, $Sr_{3}Cr_2O_8$ is an isotropic magnetic system
\cite{aczel}, unlike DTN, which has a condensed state of the
$XY$ type. In the inset of Fig.~\ref{fig5}, we show the data for the
specific heat of DTN along the QCT at $H_{C1}$. This follows
approximately a $T^{3/2}$ behavior, in agreement with our conclusion
that the purely Gaussian contribution ($u=0$) $C/T \propto T^{(d-z)/z}$
dominates the thermal behavior of the specific heat.

Finally we discuss the critical behavior of the longitudinal
susceptibility $\chi_{\|}$ of the magnetic Bose system. At zero
temperature, $\chi_{\|} \propto \partial^2 f_S/\partial H^2$ scales
as the compressibility of the Bose gas,
\begin{equation} 
\chi_{\|} \propto |g|^{-\alpha}K\left[\frac{T}{|g|^{\nu z}}\right]
\end{equation} 
Notice that, in three dimensions, the quantum Gaussian exponent $\alpha=2-\nu(d+z)=-1/2$ is
negative and the longitudinal susceptibility vanishes at the zero
temperature phase transition. Mean-field would give rise to at most a jump in the
longitudinal susceptibility at the transition. 

At $T=0$ when the external magnetic
field is larger than the upper critical field $H_{C2}$, the system is
fully polarized and the scaling function $K(t=0)=0$. As temperature
increases for $H>H_{C2}$ and below the line $T<T_G=|g|^{\nu z}$, this
scaling function can be easily guessed, $K(t)= \exp(-T_G/T)$, since
the system is gapped in this region of the phase diagram.

For finite temperatures, close to the critical line $H_C(T)$, we
obtain $\chi_{\|}$ from the second derivative of the free energy with
respect to the magnetic field.  Using Eq.~(\ref{free}), we get,
\begin{equation} \chi_{\|} \propto \frac{\partial^2 f_S}{\partial H^2}
\propto T^{\frac{\tilde{\alpha}-\alpha}{\nu z}}\left|
\frac{H-H_C(T)}{H_{C0}} \right|^{-\tilde{\alpha}}
\end{equation} 
Along the quantum critical trajectory, then, we get
$\chi_{\|} \propto T^{(\tilde{\alpha}-\alpha)/\nu
z}(uT^{1/\psi})^{-\tilde{\alpha}}$. For DTN we expect to find
$\chi_{\|} \propto \sqrt{T} \ln T$ along the QCT, since
$\tilde{\alpha} \approx 0$.

\section{Discussion}

Although discussed in the context of insulating magnetic systems
presenting a Bose-Einstein condensation of magnons, the results in this
paper should apply to any quantum phase transition. In particular,
they also describe quantum criticality in metallic systems. For
example, the equation for the amplitude of the specific heat at a
thermal phase transition near a QCP,
$$
C \propto T^{\frac{5}{2}}
$$
holds near any superconductor quantum critical point with dynamic
exponent $z=2$ with the thermal superconductor transition in the
universality class of the three-dimensional $XY$ model.

The present approach relies only on universal features of continuous
quantum phase transitions: we only assume that, near the transition, a
characteristic length and a characteristic time contain all relevant
information on the system. Any difference in the nature of the
materials, whether metal or insulator, or in the phase transition
itself, whether magnetic or superfluid, will be accounted for by
the various critical exponents, in special by the dynamic
exponent $z$, which distinguish different universality classes.

\section{Conclusion}

We have used a generalized scaling approach to study the critical
behavior of physical quantities near a quantum phase transition. We
considered systems above their lower critical dimension, such that a
line of finite-temperature continuous phase transitions emanates from
the QCP. Our approach has shown that for quantum phase transitions
described by Gaussian exponents, i.e., with $d+z >d_C$, where $d_C$ is
the upper critical dimension, classical and quantum fluctuations are
intermingled near the QCP. The breakdown of the generalized scaling
relation $\psi=\nu z$, between the shift and the crossover exponents,
signals the intertwining. In general this breakdown is due to a
dangerous irrelevant interaction on the Gaussian fixed point
describing the zero-temperature phase transition.  

Our approach brings into the stage of quantum criticality, the thermal
phase transitions themselves. We find that thermal critical
fluctuations close to a QCP are suppressed by quantum effects and
their study allows a deeper understanding of quantum criticality. As
future experiments progress towards lower temperatures, more detailed
comparisons with theory will become possible allowing finer
distinction between alternative scenarios.

We have shown that the momentum space RG and the self-consistent
approach yield quantum-criticality results that follow from the 
GSA under the assumption of Gaussian exponents for the thermal phase
transitions. Whether this is a sufficiently good description of the
thermal phase transitions and, as implied by our analysis, of
quantum criticality itself can be tested experimentally. The
alternative assumption, that the thermal exponents have the expected
Wilson character, has stronger experimental consequences. The
amplitude of critical thermal fluctuations will be reduced and the
physical behavior along the quantum critical line will show deviations
from the naively expected quantum critical exponents.

We found the behavior of the specific heat near the QCP to be
different from that of other physical quantities such as, for example,
the order parameter susceptibility or the correlation length. The
former has its amplitude suppressed near the QCP even assuming
Gaussian thermal exponents. Besides, the corrections for this quantity
due to the irrelevant interaction can safely be neglected when
compared to the, in general more singular, purely Gaussian ($u=0$)
result. This is definitely not the case for other physical quantities
where the corrections in $u$ to the purely Gaussian result have $u$ in
the denominator due to its dangerously irrelevant character.

\section{Acknowledgements}

I would like to thank Armando Paduan-Filho for useful
discussions. I would like also to express my gratitude for the referee
for his suggestions and advices concerning the historical aspects of the development
of the theory of quantum phase transitions. Work partially supported by Conselho Nacional de
Desenvolvimento Cient\'{\i}fico e Tecnol\'{o}gico - CNPq and
Funda\c{c}\~ao de Amparo a Pesquisa do Estado do Rio de Janeiro -
FAPERJ.

\end{document}